# THE RELATIONAL DATABASE ASPECTS OF ARGONNE'S ATLAS CONTROL SYSTEM

D. E. R. Quock, F. H. Munson, K. J. Eder[*], S. L. Dean[*], ANL, Argonne, IL 60439, USA

Abstract

Argonne's ATLAS (Argonne Tandem-Linac Accelerator System) control system comprises two separate database concepts. The first is the distributed real-time database structure provided by the commercial product Vsystem [1]. The second is a more static relational database archiving system designed by ATLAS personnel using Oracle Rdb [2] and Corel Paradox [3] software. The configuration of the ATLAS facility has presented a unique opportunity to construct a control system relational database that is capable of storing and retrieving complete archived tune-up configurations for the entire accelerator. This capability has been a major factor in allowing the facility to adhere to a rigorous operating schedule. Most recently, a Web-based operator interface to the control system's Oracle Rdb database has been installed. This paper explains the history of the ATLAS database systems, how they interact with each other, the design of the new Web-based operator interface, and future plans.

## 1 ATLAS CONTROL SYSTEM

ATLAS is a linear accelerator system containing 3 separate ion source injectors and 64 superconducting resonators. This facility accelerates low-energy, heavy-ion elements ranging from hydrogen to uranium. These ions are accelerated up to a maxiumum of 20% the speed of light mostly for the purpose of nuclear physics research. Typical operation at ATLAS is to start a new experiment every three to five days, with beam being delivered to one of three target areas 24 hours a day.

### 1.1 Real-time Control

Real-time control of the accelerator is achieved by the interface of Vista's Vsystem software running on a Compaq AlphaServer [4] computer to a CAMAC (Computer Automated Measurement And Control) Serial Highway. Vsystem graphical user interface displays are provided on workstations located throughout the accelerator facility for operator real-time control and readback of accelerator devices.

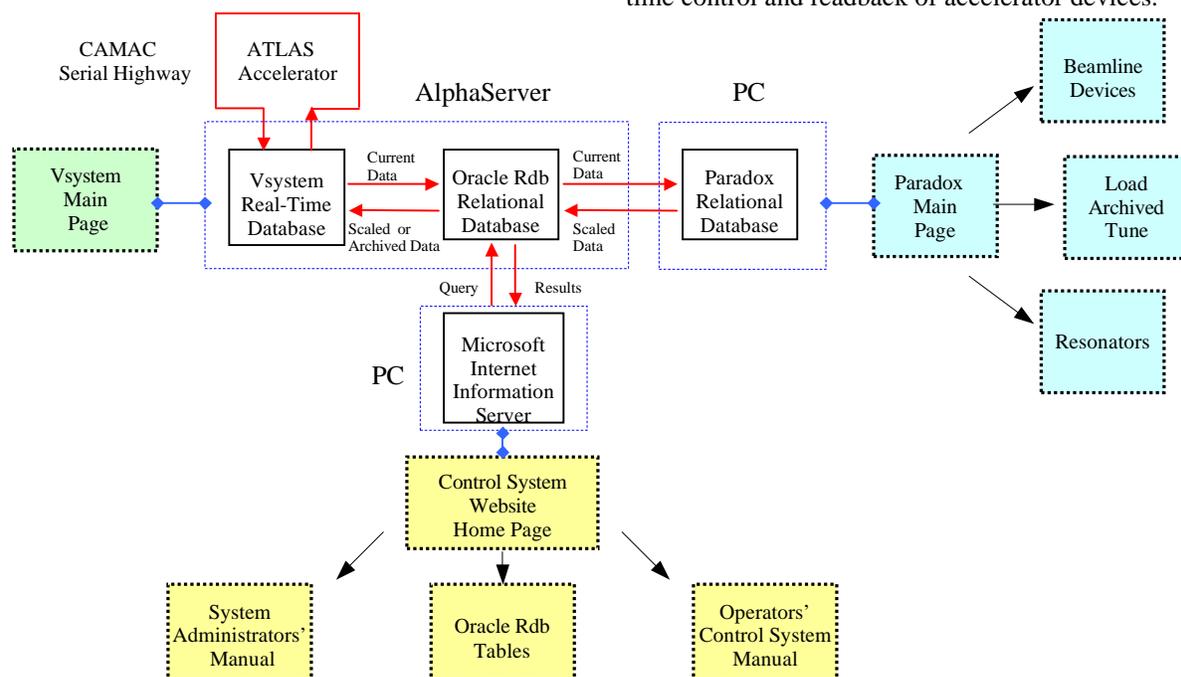

Figure 1: ATLAS Database Management System Software Configuration

---

[*] Undergraduate Research Participants

# 2 ATLAS DATABASE SYSTEMS

## 2.1 Historical Development

The ATLAS Control System real-time database, Vaccess, is augmented by two relational database management systems: Oracle Rdb and Corel Paradox. A relational database is a collection of tables connected in a series of relationships that, in the case of ATLAS, reflects the organization of accelerator devices and control system processes. The Oracle Rdb relational database was initially developed to store information about the accelerator that changed very seldom. In-house written processes retrieve information from the Oracle Rdb tables for specific functions that they perform as part of the accelerator control system. An example of this is a process that retrieves predefined stepper motor operating positions for a specific slit, foil, or other beamline insertion device. Another in-house written process continuously scans Vsystem's real-time database, and writes critical values to Oracle Rdb database files for data archiving and acceleration configuration restoration purposes. Examples of tables stored in the Oracle Rdb database are:

- Resonators
- Beam Measurement
- Cryogenic Alarms
- CAMAC Crates and Modules

With the later addition of the Paradox tune archiving database application, some of the Oracle Rdb tables began to also serve as an intermediate storage location for the transfer of data from and to Vsystem and Paradox.

## 2.2 Tune Archiving System

To expedite the startup of new ATLAS experiments, a tune archiving relational database system was developed. This system provides a means for reconfiguring accelerator components by scaling component values of a previously archived experiment according to the charge states, energies and mass requirements of the new experiment. The scaled values are then loaded into the Vsystem real-time database.

At the start of the archiving system development work, Oracle's graphical user interface software was too costly. Consequently, Corel's PC-based Paradox software was selected. Since its inception in 1994, the ATLAS accelerator-tune archiving system has expanded to include several other significant operator functions. The archiving system records accelerator operating parameters on an automated 4-hour schedule, and provides a means for monitoring and analyzing the performance of an ongoing experiment. Details of the Paradox tune archiving database application can be found in a previous paper [5].

# 3 WEB-BASED USER INTERFACE

## 3.1 ATLAS Control System Website

The most recent software application addition to the ATLAS Control System is the ATLAS Control System website. This website contains the following three major utilities:

- Interactive query of the Oracle Rdb database
- Operators' Control System Manual
- System Administrators' Manual

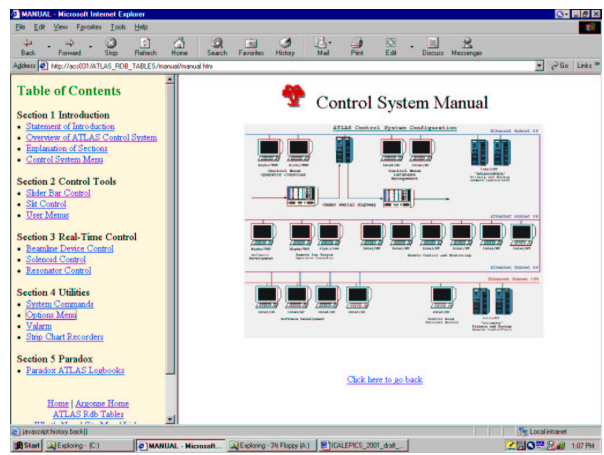

Figure 2: Online Operators' Control System Manual

The Web server software for the ATLAS Control System website is Microsoft IIS (Internet Information Server) [6]. Microsoft IIS is part of the Windows NT 4.0 Option Pack, and is integrated with the Windows NT Server operating system. The control system web pages were developed with HTML, JavaScript, and Microsoft ASP (Active Server Pages) languages. ODBC (Open Database Connectivity) is used to establish communication with the Oracle Rdb database that is installed on a Compaq AlphaServer computer. The Web server PC, AlphaServer computer, and Paradox PC reside on an isolated Ethernet local area network (LAN), and use the TCP/IP protocol for transferring information. Currently, eight PCs are available to ATLAS personnel for accessing the control system website.

## 3.2 Online Database Access

In the past, control system developers used interactive SQL (Structured Query Language) on the AlphaServer's OpenVMS operating system to view the contents of the Oracle Rdb database. This method is cumbersome, and requires knowledge of the database structure and SQL language. An Internet browser-based user interface to the Oracle Rdb database was created to provide a quick and simple querying procedure.

## 3.3 Methodology of Web Browser Query

Standard HTML pages and Microsoft ASP files provide query access to the Oracle Rdb database. The HTML pages use HTML forms that include check boxes, text entry, and radio buttons. These graphical tools provide the interactive interface to the user. The ASP files contain server-side script commands that build an SQL search string based on the user's selections and input. After a user submits a query request from the HTML web page, the ASP script commands execute on the Microsoft IIS Web server. The results of the query are returned in tabular format as an HTML page to the client browser. As shown in Figure 4, the interactive query web pages consist of two frames. The top frame allows filtering and sorting, while the lower frame displays the query results.

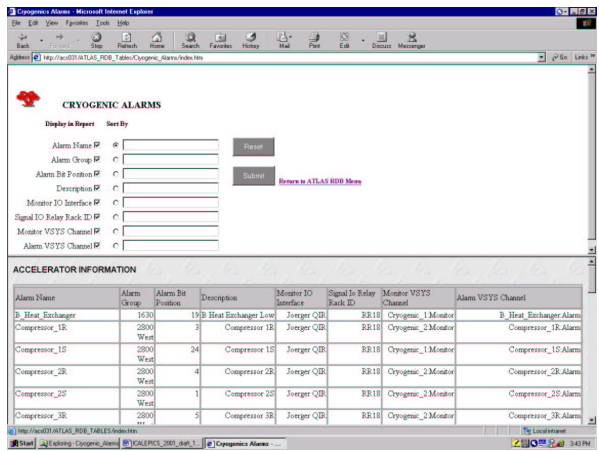

Figure 3: Oracle Rdb Interactive Query Web Page

## 4 CONCLUSION

The core of the ATLAS Control System database system is Vsystem's Vaccess real-time database. The two relational database applications, built with Oracle Rdb and Corel Paradox, are invaluable accelerator configuration and restoration utilities. The new Web-based user interface to the Oracle Rdb database has become a convenient and frequently used diagnostic tool. The online control system manual is easily accessed from remote PCs located throughout the accelerator facility.

Future plans will shift focus away from the databases themselves to improving the various interfaces to the databases. One improvement is to add a context-sensitive help system to the ATLAS Control System website. Ongoing maintenance of the website requires the update and addition of new control system documentation. It is planned to eventually open the website to TCP/IP networks outside of the ATLAS Control System network, but confined to inside Argonne National Laboratory.

This work is supported by the U.S. Dept. of Energy, Nuclear Physics Div., under contract W-31-109-ENG-38.